\documentstyle[epsfig]{article}
\begin{document} 

\def\la{\mathrel{\mathpalette\fun <}}
\def\ga{\mathrel{\mathpalette\fun >}}
\def\fun#1#2{\lower3.6pt\vbox{\baselineskip0pt\lineskip.9pt
\ialign{$\mathsurround=0pt#1\hfil##\hfil$\crcr#2\crcr\sim\crcr}}} 

\begin{center}
{\Large \bf Azimuthal anisotropy of jet quenching at LHC~\footnote{Talk given at 
4th International Conference "Physics and Astrophysics of Quark-Gluon Plasma", 
November 26-30, 2001.}} \\

\vspace{4mm}

\underline{I.P.~Lokhtin}, S.V.~Petrushanko, L.I.~Sarycheva and A.M.~Snigirev  \\
M.V.Lomonosov Moscow State University, D.V.Skobeltsyn Institute of Nuclear Physics \\
119899, Vorobievy Gory, Moscow, Russia\\
\end{center}  

\begin{abstract} 
We analyze the azimuthal anisotropy of jet spectra due to energy loss of hard partons in 
quark-gluon plasma, created initially in nuclear overlap zone in collisions with 
non-zero impact parameter. The calculations are performed for 
semi-central Pb$-$Pb collisions at LHC energy. 
\end{abstract}

\section {Introduction}   

High-$p_T$ jet production and other hard processes are considered 
as a one of the most promising tools for studying properties of super-dense and 
hot matter created in nucleus-nucleus collisions at RHIC and LHC. The challenging 
problem here is the behaviour of colour charge in quark-gluon matter 
associated with the coherence pattern of the medium-induced radiation, 
resulting in a number of interesting non-linear phenomena, in particular, the 
dependence of radiative energy loss per unit distance $dE/dx$ along the total 
distance traversed (see review~\cite{baier_rev} and references therein).  
In our previous work~\cite{lokhtin00} we predicted that medium-induced parton 
rescattering and energy loss should result in a dramatic change in the 
distribution of jets over impact parameter as compared to what is expected from 
independent nucleon-nucleon interactions pattern. In this paper we concentrate on the 
phenomena related to the azimuthal dependence of jet energy loss and corresponding jet 
spectra in semi-central heavy ion collisions at LHC energies, when the cross section
for hard jet production at $E_T \sim 100$ GeV scale is large enough to study the impact 
parameter dependence of such processes. We consider the experimental conditions of CMS
experiment at LHC~\cite{cms94}, which can provide jet reconstruction and adequate 
measurement of impact parameter of nuclear collision using calorimetric 
information~\cite{note00-060}. Note that the possible azimuthal anisotropy of
high-$p_T$ hadron spectra at RHIC was discussed in a number of 
papers~\cite{uzhi,wang00,gyul00}.

\section {Nuclear geometry and jet energy loss} 

The details of geometrical model of jet production and jet passing through a dense 
matter in high energy symmetric nucleus-nucleus collision can be found 
in~\cite{lokhtin00}. The figure 1 shows the essence of the problem in the 
plane of impact parameter {\bf b} of two colliding nuclei $A$-$A$. 
The initial distribution over jet production vertex 
$B (r, \psi)$ in nuclear overlap zone at given impact parameter $b$ is written as 
\begin{equation} 
\label{vertex}
P_{AA}({\bf r}, b) = \frac{T_A(r_1)\cdot T_A(r_2)}{T_{AA}(b)}, 
\end{equation}
where ${\bf r}=r\cos{\psi}\cdot{\bf e_x}+r\sin{\psi}\cdot{\bf e_y}$ is the vector from
beam axis $z$ to vertex $B$;   
$r_{1,2} = \sqrt{r^2+b^2/4 \pm rb\cos \psi}$ 
is the distance between nucleus centers ($O_1, O_2$) and vertex $B$; $T_{AA}(b)$
and $T_A({\bf r})$ are the standard nuclear overlap and nuclear thickness functions
respectively. 

\begin{figure}[htb]
\begin{minipage}[t]{82mm}
\resizebox{82mm}{60mm} 
{\includegraphics{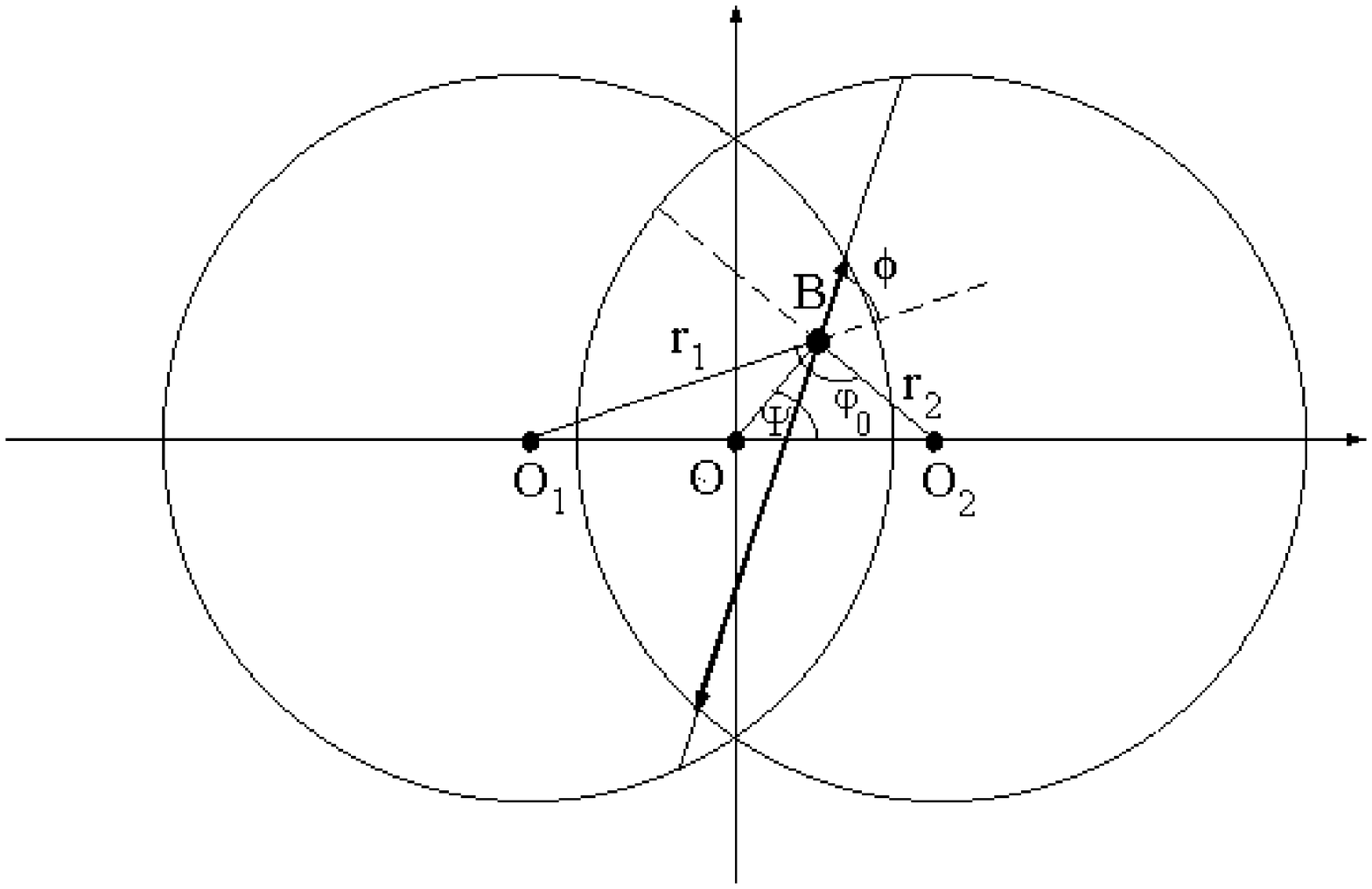}} 
\caption{\small Jet production in high energy symmetric nucleus-nucleus collision in 
the plane of impact parameter {\bf b}. $O_1$ and $O_2$ are nucleus centers, 
$OO_2 = -O_1O = b/2$. $B$ is dijet production 
vertex; $r$ is the distance from the beam axis to $B$; $r_1,r_2$ are 
distances between $O_1$, $O_2$ and $B$.} 
\label{fig:1}
\end{minipage}
\hspace{\fill}
\begin{minipage}[t]{75mm}
\resizebox{75mm}{75mm}  
{\includegraphics{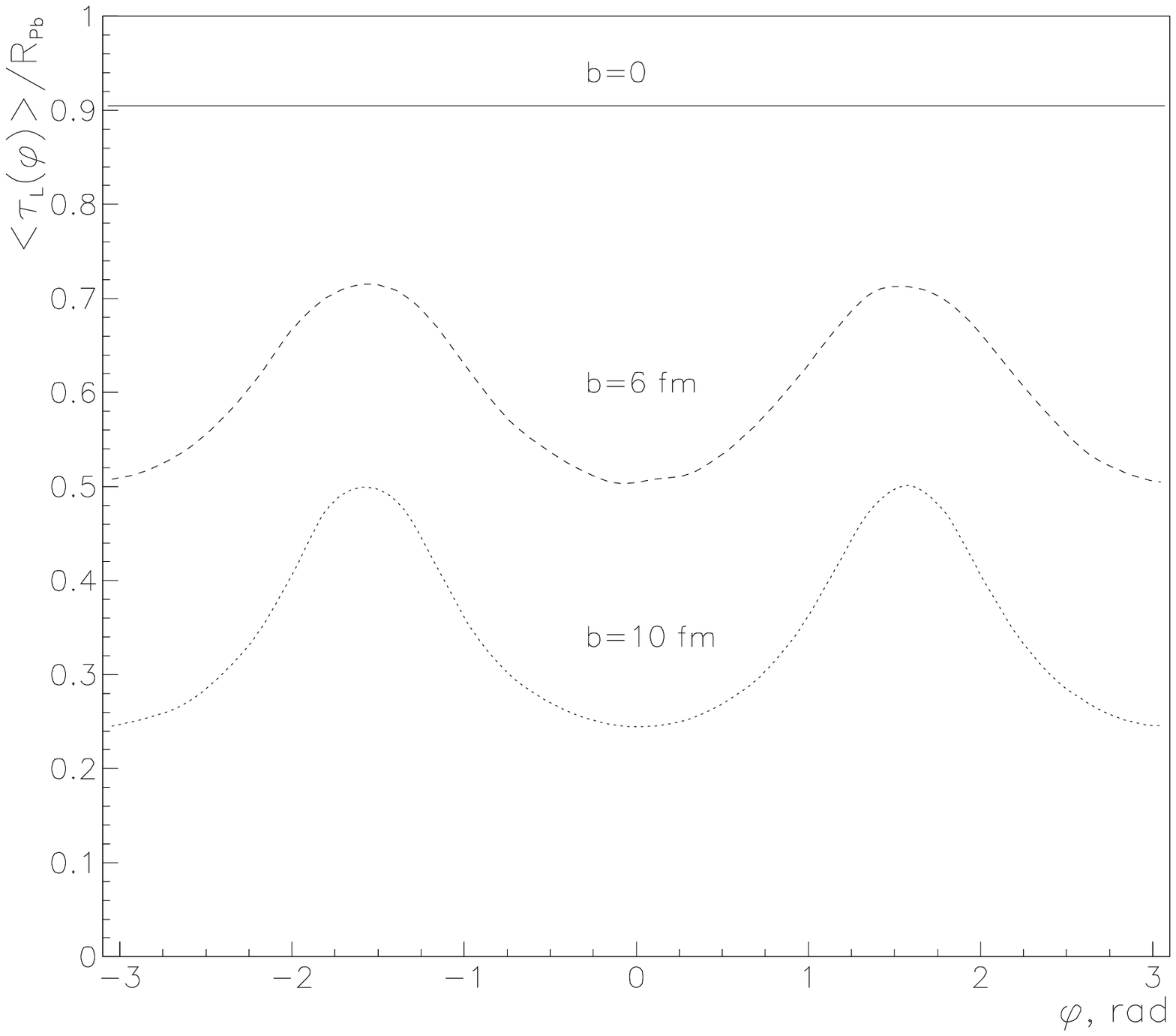}} 
\caption{\small Average proper time $\tau_L/R_A$ of jet escaping from the dense matter 
as a function of jet azimuthal angle $\varphi$ relatively the nuclear reaction
plane.  The curves (from top to bottom) correspond to the impact parameter 
values $b = 0$, $6$ and $10$ fm.}  
\label{fig:2}
\end{minipage}
\end{figure}

The basic kinetic integral equation for the energy loss $\Delta E$ as a function of 
initial energy $E$ and path length $L$ has the form 
\begin{eqnarray} 
\label{elos_kin}
\Delta E (L,E) = \int\limits_0^Ldx\frac{dp(x)}{dx}
\lambda(x)\frac{dE(x,E)}{dx} \, , ~~~~ 
\frac{dp(x)}{dx} = \frac{1}{\lambda(x)}\exp{\left( -x/\lambda(x)\right) }
\, ,  
\end{eqnarray} 
where $x$ is the current transverse coordinate of a parton, $dp/dx$ is the scattering
probability density, $dE/dx$ is the energy loss per unit length, $\lambda = 1/(\sigma 
\rho)$ is in-medium mean free path, $\rho \propto T^3$ is medium density at temperature 
$T$, $\sigma$ is the integral cross section of parton interaction in the medium.
It is straightforward to evaluate the time $\tau_L = L$ it takes for jet to traverse 
the dense zone:   
\begin{equation} 
\label{taul} 
\tau_L = min\{\sqrt{R_A^2 -r_1^2\sin^2\phi} - r_1
\cos\phi,~\sqrt{R_A^2 - r_2^2 \sin^2(\phi-\varphi_0)} - r_2
\cos(\phi-\varphi_0)\} ,  
\end{equation} 
where $\phi = \varphi - (\psi/|\psi|) \arccos{\{ (r\cos{\psi}+b/2)/r_1)\} }$ is the 
isotropically distributed  angle which determines the direction of a jet 
relatively to vector ${\bf r_1}$, $\varphi$ is the azimuthal angle between the 
direction of a jet motion and vector ${\bf b}$, $\varphi_0 = (\psi/|\psi|) \arccos
{(r^2-b^2/4)/(r_1r_2)}$ is angle between vectors ${\bf r_1}$ and ${\bf r_2}$. 
One can see from eq.(\ref{taul}) that for non-central collisions, $b \ne 0$, value 
$\tau_L$ depends on $\varphi$: it is maximum at $\varphi = \pm \pi/2$ and minimum 
at $\varphi = 0$ (see fig.2 for Pb$-$Pb collisions and impact 
parameters values $b = 0$, $6$ and $10$ fm). Since energy loss is increasing 
function of jet in-medium path-length, it will then depends on $\varphi$ also. 

In order to illustrate the azimuthal anisotropy of parton energy loss, we treat the 
medium as a boost-invariant longitudinally expanding quark-gluon fluid, and partons as 
being produced on a hyper-surface of equal proper times $\tau = \sqrt{t^2 - 
z^2}$~\cite{bjorken}. For certainty we used the initial conditions for the 
gluon-dominated plasma formation 
expected for central $Pb-Pb$ collisions at LHC~\cite{eskola94}: 
$\tau_0 \simeq 0.1$ fm/c, $T_0 \simeq 1$ GeV, $N_f \approx 0$, $\rho_g \approx 
1.95T^3$. For non-central collisions we suggest the proportionality of the initial 
energy density $\varepsilon_0$ to the ratio of nuclear overlap function $T_{AA}(b)$
and effective transverse area $S_{AA}(b)$ of nuclear overlapping, 
$\varepsilon _0 (b) \propto T_{AA}(b)/S_{AA}(b)$~\cite{lokhtin00}. 

Our approach relies on an accumulative energy losses, when gluon radiation is 
associated with each scattering in expanding 
medium together including the interference effect by the modified radiation spectrum as 
a function of decreasing temperature $dE/dx(T)$. For our calculations we have used
collisional part of loss and differential scattering cross section from our 
work~\cite{lokhtin00}; the energy spectrum of coherent medium-induced gluon radiation 
was estimated using BDMS formalism~\cite{baier}. It is important to notice that 
the coherent LPM radiation induces a strong dependence of the jet energy on the jet 
cone size~\cite{baier,lokhtin98,urs,vitev}, while the 
collisional energy loss turns out to be practically independent on cone size, because 
the bulk of "thermal" particles 
knocked out of the dense matter by elastic scatterings fly away in almost 
transverse direction relative to the jet axis~\cite{lokhtin98}. 

\section{Azimuthal anisotropy of jet spectra} 

Figure 3 shows the average value of medium-induced radiative (a) and 
collisional (b) energy loss of quark with initial transverse energy $E_T^q = 100$ 
GeV as a function of $\varphi$. As it might be expected, azimuthal anisotropy of 
energy loss goes up with increasing $b$, because  azimuthal asymmetry of the 
volume gets stronger in this case. On the other hand, the absolute value of 
energy loss, of course, goes down with increasing $b$ due to reducing 
absolute value of mean distance traversed (and also due to decreasing initial 
energy density of the medium at $b \ga R_A$). Then the non-uniform 
dependence of energy loss on azimuthal angle results in azimuthal anisotropy 
of jet spectra in semi-central collisions. 
Figure 4 shows the distribution of jets over azimuthal angle $\varphi =\varphi_{1,2}$ 
for the cases with collisional and radiative loss (a) and collisional loss only (b) for 
$b = 0$, $6$ and $10$ fm (the initial jet distributions have been generated using 
PYTHA$\_5.7$ model~\cite{pythia}). The CMS kinematical acceptance for jets was taken 
into account: $E_T^{jet} > 100$ GeV, $|y^{jet}| < 2.5$. 
The distributions are normalized on the initial distributions of jets over
$\varphi$ in Pb$-$Pb collisions (without any energy loss). We can see that
the azimuthal anisotropy gets stronger as going from central to semi-central
collisions, but the absolute suppression factor reduces with increasing $b$. 
For jets with finite cone size one can expect the intermediate result between
cases (a) and (b), because, as we have mentioned before, radiative loss
dominates at relatively small angular sizes of jet cone $\theta_0 (\rightarrow 0)$, 
while the relative contribution of collisional loss grows with increasing
$\theta_0$. 

\begin{figure}[htb]
\begin{minipage}[t]{78mm}
\resizebox{78mm}{78mm} 
{\includegraphics{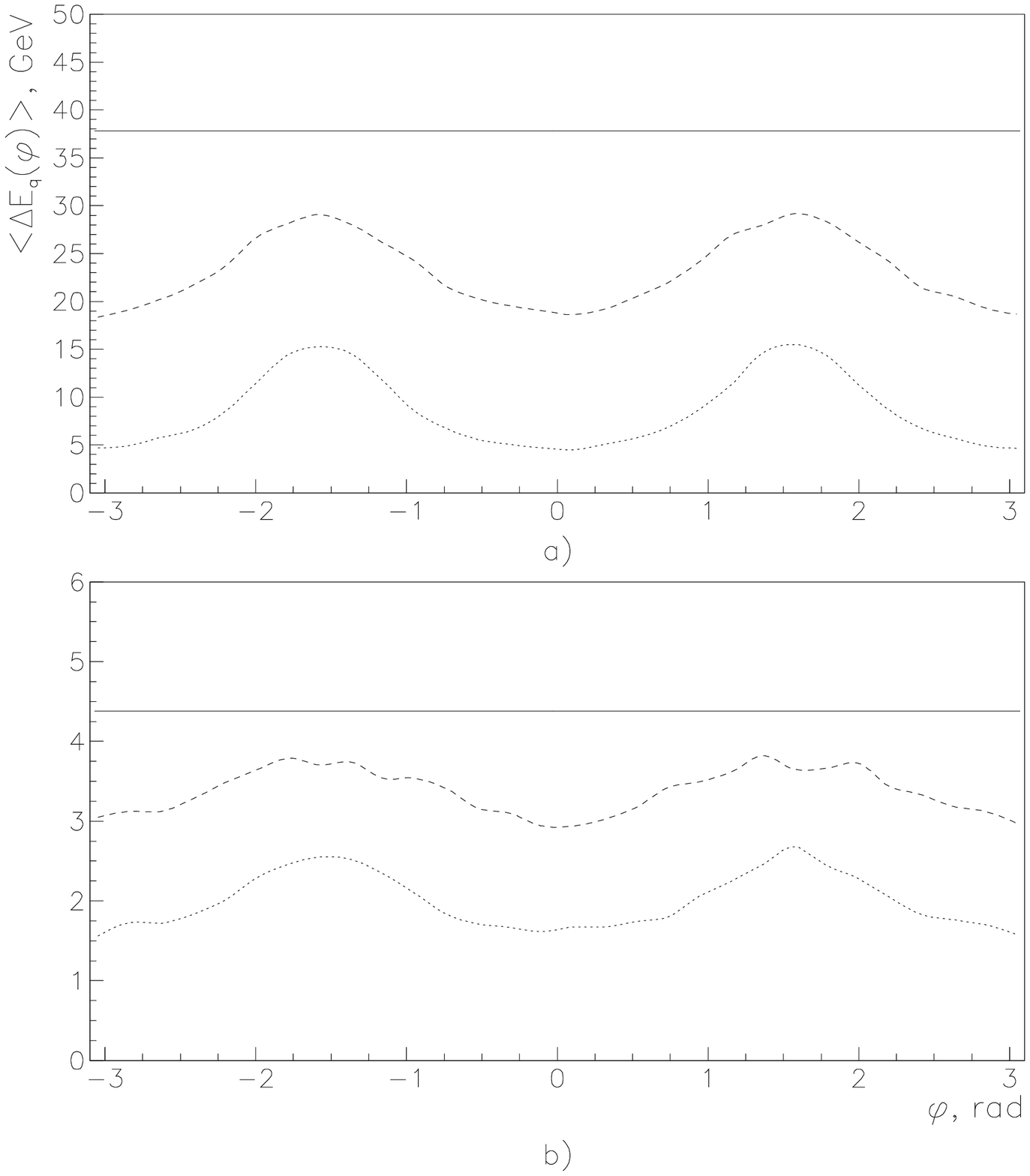}} 
\caption{\small The average medium-induced radiative (a) and collisional (b) energy 
loss of quark with initial transverse energy $E_T^q = 100$  GeV as a function of 
quark azimuthal angle $\varphi$. The curves (from top to bottom) correspond to 
the impact parameter values $b = 0$, $6$ and $10$ fm.}  
\label{fig:3}
\end{minipage}
\hspace{\fill}
\begin{minipage}[t]{78mm}
\resizebox{78mm}{78mm}  
{\includegraphics{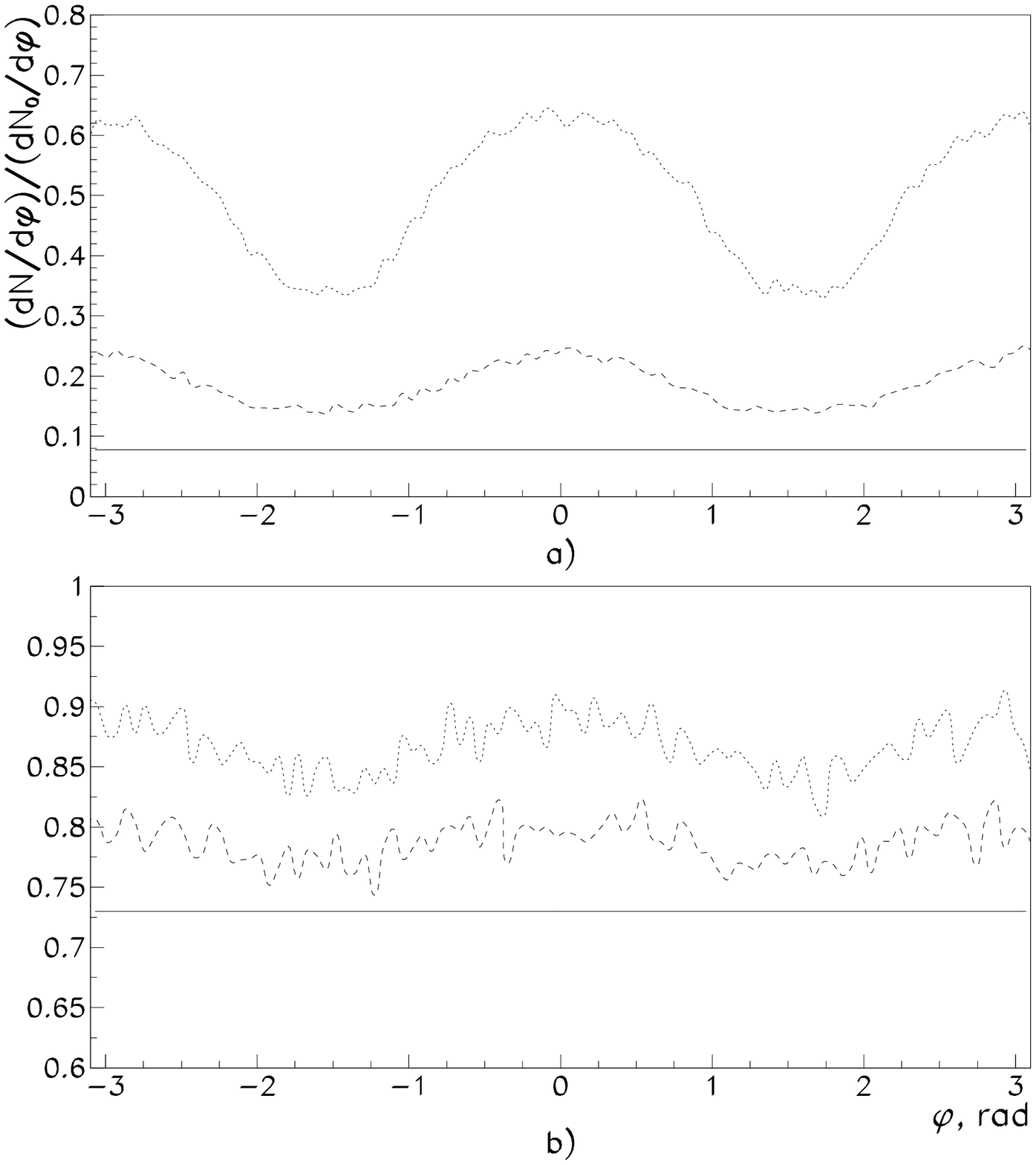}} 
\caption{\small The distribution of jets over azimuthal angle for the cases with 
collisional and radiative loss (a) and collisional loss only (b), jet
kinematical acceptance is $E_T^{jet} > 100$ GeV and $|y^{jet}| < 2.5$. 
The histograms (from bottom to top) correspond to the impact parameter 
values $b = 0$, $6$ and $10$ fm.}  
\label{fig:4}
\end{minipage}
\end{figure}

In non-central collisions the jet distribution over $\varphi$ is approximated well
by the following form,  
$dN/d \varphi = A (1+B\cos{2 \varphi})~$,  
where $A=0.5(N_{max}+N_{min})$ and $B=(N_{max}-N_{min})/
(N_{max}+N_{min})=2\left< \cos{2\varphi}\right> $. The average cosines of  
$2 \varphi$ for particle flow is called as coefficient of azimuthal 
anisotropy $v_2$~\cite{voloshin}. In our model the coefficient of jet azimuthal 
anisotropy increases almost linearly with growth of $b$ and becomes maximum at 
$b \sim 1.2 R_A$, after that it reduces rapidly with increasing $b$
(the domain of impact parameter values, where the effect of decreasing energy loss 
due to reducing effective transverse size of dense zone and initial energy density 
of the medium is crucial and not compensated anymore by  
stronger non-symmetry of the volume). Other important feature is the jet azimuthal 
anisotropy decreases with increasing jet energy, because the energy dependence
of medium-induced loss is rather weak~\cite{baier,urs}. 

In conclusion of this section we remark, that the methodical advantage of 
azimuthal jet observables is obvious: one needs to reconstruct only azimuthal 
position of jet, but not the total jet energy. It can be done more easily and with
high accuracy, while the reconstruction of the jet energy is more ambiguous
task~\cite{note00-060}. On the other hand, the performance of inclusive analysis of jet 
production as a function of azimuthal angle requires event-by-event determination of 
the reaction plane angle. Summarized in papers~\cite{voloshin} present methods of 
determination of the reaction plane angle are applied to study elliptic flow of soft 
particles in current heavy ion dedicated experiments at SPS 
and RHIC. The capability of CMS tracker to reconstruct momenta of all 
semi-hard ($p_t\ga 2$GeV$/c$), and especially soft ($p_t\la 2$GeV$/c$) charged
particles is not clear at the moment. 
However the transverse energy flow in central CMS calorimeters 
should reflect the pattern of semi-hard particles flow, in particular, including any 
azimuthal anisotropy manifestation. Thus the determination of nuclear reaction 
plane angle using semi-hard particles flow (not incorporated in high-$p_T$ jet 
pair) could be, in principle, possible due to two factors:  
({\it 1}) sensitivity of semi-hard particles to the azimuthal asymmetry of reaction
volume under condition that the most part of them being the products of 
in-medium radiated gluons~\cite{uzhi,wang00}; 
({\it 2}) predicted high enough multiplicity of such particles at LHC energies 
(which is comparable, for example, with the total multiplicity at SPS). 

\section{Conclusions} 

The interesting phenomenon is predicted to be observed in semi-central heavy ion 
collisions at LHC: the appearance of azimuthal anisotropy of jet spectra due to 
energy loss of jet partons in azimuthally non-symmetric volume of dense 
quark-gluon matter, created initially in nuclear overlap zone. We have found that the 
coefficient of jet azimuthal anisotropy increases 
almost linearly with growth of $b$ up to $b \sim 1.2 R_A$ fm. 
The effect of jet azimuthal anisotropy decreases slightly with jet energy.  

The methodical advantage of azimuthal jet observables is that one needs to 
reconstruct only azimuthal position of jet, but not the total jet energy. On the
other hand, the performance of inclusive analysis of jet production as a function 
of azimuthal angle requires event-by-event determination of the reaction plane 
angle. We suggest that under LHC conditions the existing methods of determination 
of nuclear reaction plane angle might be applied with measuring the azimuthal 
anisotropy of global transverse energy flow originated from mini-jet production in 
non-symmetric volume of dense QCD-medium.

{\it Acknowledgements} 
Discussions with M.~Bedjidian, D.~Denegri, P.~Filip, V.~Uzhinskii and U.~Wiedemann 
are gratefully acknowledged. I.~Lokhtin thanks the organizers of Fourth 
International Conference "Physics and Astrophysics of Quark-Gluon Plasma" for the warm 
welcome and stimulating atmosphere.

\end{document}